# Passage of neutrons through accelerating crystal and the acceleration effect

A.I. Frank[1], V.A. Bushuev[2], M.A. Zakharov[1] and G.V. Kulin[1]

[1]*Joint Institute for Nuclear Research, 141980 Dubna, Russia*
[2]*Moscow State University, 119991 Moscow, Russia*

**Abstract.** The problem of changing the energy of a neutron when it passes through an accelerating crystal under conditions close to the Bragg condition is considered. It is shown that, similar to the case of the passage of long-wavelength neutrons through a refractive sample, the accelerated motion of a crystal results in a change in neutron energy. The physical nature of the phenomenon in both cases is determined by the difference in the Doppler frequency shift at the entrance and exit of the wave through the sample. The difference between the case of ordinary refraction of long-wavelength neutrons and the case of wave passage through a crystal under conditions where diffraction has a dominant influence on the nature of wave propagation is only quantitative. The results obtained are qualitatively consistent with the available experimental data.

**1. Introduction.**

In [1-2], an effect was observed consisting in a change in the energy of neutrons as they pass through a refractive sample moving with acceleration. A similar effect was previously predicted for light [3]. The qualitative explanation of these two phenomena was reduced to the idea of the differential Doppler effect. Since the velocity of the sample changes during the propagation of the wave through it, the magnitude of the Doppler frequency shifts at refraction at the input and output surfaces of the sample differs not only in sign, but also in absolute value. This interpretation of the effect is completely valid in the case of neutron wave refraction, where physically significant phenomena occur only at the boundaries of the sample, and the fact of its movement does not affect the nature of wave propagation through matter. But it is hardly entirely correct for light [4].

A universal approach to the description of optical phenomena that occur when a wave passes through a refractive sample moving with acceleration was formulated in [1]. It has been shown that, based on the equivalence principle alone, it is possible to describe the change in frequency and energy in a single manner for both neutrons and light when they pass through a refractive medium. In this case, the main factor determining the magnitude of the frequency shift when a wave passes through a sample moving with constant acceleration, is the amount of time delay τ due to the difference the velocities of neutrons in matter and vacuum. As a result, the concept of a universal optical phenomenon emerged, valid for waves of any nature and was called in the work [1] the accelerating matter effect (AME). It was subsequently realized that the idea of the effect being related only to refraction was unjustifiably narrow, and that a change in the frequency of the wave would inevitably occur when interacting with any object moving with acceleration [5]. The frequency change in all cases is determined by the ratio

$$\Delta\omega = k_0 a \tau, \tag{1}$$

where $k_0$ is the wave number of the incoming wave and $a$ is acceleration.

The time delay between the scattered and the incident waves can occur for various reasons. In neutron optics, the occurrence of a time delay that is unrelated to ordinary refraction has been demonstrated in experiments in which neutrons passed through a crystal under conditions close to the Bragg condition [6-8].

From relation (1) it follows that setting the sample into accelerated motion under these conditions should also lead to a change in the energy of the neutrons passing through it. In order to determine the magnitude of the change in frequency and energy it is only necessary to correctly calculate the magnitude of the time delay resulting from the difference in the group velocity of the neutron in the crystal from its vacuum value.

**2. The neutron velocity in a crystal under conditions close to the Bragg condition.**

The starting point for analyzing the corresponding problem is the concept of the effective mass of a neutron in a medium characterizing an arbitrary dispersion law [9, 10]. In general, the effective mass is a tensor quantity

$$\left(\frac{1}{m^*}\right)_{\mu\nu} = \frac{1}{\hbar}\frac{\partial^2}{\partial k_\mu \partial k_\nu}\omega(\mathbf{k}), \tag{2}$$

where $\mathbf{k}$ is the wave vector inside the crystal.

For simplicity, we will assume below that the reciprocal lattice vector $\mathbf{g}$ of the crystal is collinear to the wave vector $\mathbf{k}_0$ of the neutron wave incident on it. In this case, we can ignore the tensor nature of the effective mass and put it equal to

$$m^* = 2mk\frac{dF}{d(k_0^2)}, \tag{3}$$

where $m$ is the true mass of the neutron, $F$ is the dispersion function that relates the wave number in the medium $k$ to the wave number of the incident wave $k_0$:

$$k = F(k_0^2). \tag{4}$$

Neutron velocity in the matter is obviously [11],

$$v = \frac{\hbar k}{m^*} = \frac{\hbar}{2m}\left[\frac{dF}{d(k_0^2)}\right]^{-1}. \tag{5}$$

It is easy to see that the same expression can be derived from the definition of group velocity in the medium $v_{gr} = \dfrac{d\omega}{dk}$. Thus, to calculate the delay time of a wave passing through a crystal, it is necessary to establish the dispersion law of waves propagating in a crystal near the Bragg conditions.

From the solution of the Schrodinger equation for a medium with periodic potential

$$U(\mathbf{r}) = \frac{\hbar^2}{2m}\chi(\mathbf{r}), \qquad (6)$$

where $\chi(\mathbf{r}) = 4\pi N(\mathbf{r})b$, $N(\mathbf{r})$ is the density of nuclei and $b$ is the neutron coherent scattering length, it can be obtained that in the two-wave approximation the wave numbers of both waves in the crystal are determined by the expression

$$k_{1,2}^2 = k_0^2 - \chi_0 - \chi_g\left(\Delta \pm \sqrt{\Delta^2 - 1}\right), \qquad (7)$$

where $\chi_0$ and $\chi_{\pm g}$ are the amplitudes of the Fourier expansion of a periodic function $\chi(\mathbf{r})$:

$$\chi(\mathbf{r}) = \chi_0 + \chi_g \exp(-igz) + \chi_{-g}\exp(igz), \qquad (8)$$

the $z$ axis is oriented normally to the crystal surface and the crystallographic planes parallel to it, and

$$\Delta = \frac{\alpha - 2\chi_0}{2\chi_{eff}} \qquad (9)$$

is normalized deviation from the exact Bragg condition $\alpha=0$, where $\alpha = (2k_0 - g)g$ and $\chi_{eff} = \sqrt{\chi_g \chi_{-g}}$.

The amplitudes $\chi_0$ and $\chi_g$ in (7) and (8) correspond to the effective potentials $U_0 = (\hbar^2/2m)\chi_0$ and $U_g = (\hbar^2/2m)\chi_g$. The value of $\alpha$ becomes zero at the so-called Bragg wave number $k_B = g/2$ of the incident neutron and the Bragg energy $E_B = \hbar^2 g^2/8m$.

The plus sign in (7) corresponds to case $\Delta < -1$ and the minus sign to case $\Delta > 1$. The region $|\Delta| < 1$ corresponds to the Bragg reflection.

Note that formula (7) for the dispersion law in the Bragg diffraction geometry, as expected, differs in the sign under the square root from the expressions obtained in [9, 12] for the case of Laue geometry. However, it does not agree with the similar expression in [13, 14], where the third term has the form $\Delta/(\Delta^2 + 1)$.

Knowing the dispersion law defined by formulas (7) - (9), we can find the instantaneous velocity of a neutron in a matter according to (5). If the crystal is moving at a variable velocity then to calculate the time of flight of a neutron through it we must, if necessary, also consider that in the crystal coordinate system the wave number $k_0$ also depends on time (for more details, see the next section).

Based on relation (7), it can be shown that in the region $|\Delta| > 1$ the group velocity in the medium $v_{gr} = \dfrac{d\omega}{dk}$ is determined by the following expression:

$$v_{gr} = \frac{\hbar k_{1,2}}{m} \frac{\sqrt{\Delta^2 - 1}}{|\Delta|}. \qquad (10)$$

With a significant deviation of the energy of incident neutrons from the Bragg energy, i.e. at $|\Delta| \gg 1$, the group velocity tends to the value $v_{gr} = (\hbar k_0 / m)n = v_0 n < v_0$, which is typical for the amorphous medium, where $v_0$ is the velocity of the incident neutron in a vacuum, and $n = \sqrt{1 - \chi_0 / k_0^2}$ is the refractive index.

Figure 1 shows the result of calculation the neutron velocity near the Bragg conditions for the (110) plane of a quartz crystal. The parameters used were the values given in the article [13], namely $U_0 = 10^{-7}\,\text{eV}$, $U_g = 4 \times 10^{-8}\,\text{eV}$, $E_B = 3.4 \times 10^{-3}\,\text{eV}$.

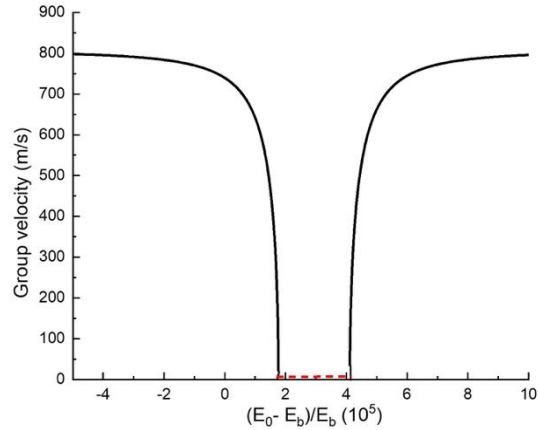

Fig. 1. The group velocity of a neutron in a quartz crystal near the Bragg conditions as a function of neutron energy. The red dotted line corresponds to the region of Bragg reflection, where there is no transmitted wave.

It can be seen from the figure that in the vicinity of the Bragg reflection region the neutron velocity in a crystal strongly depends on energy and significantly differs from the neutron velocity in vacuum, which is on the order of 800 m/s in this case. Obviously, the time required for a neutron to pass through a crystal under these conditions should also increase. Such an increase in the time of flight through the crystal when approaching Bragg's conditions was observed in [8].

**3. The change in neutron energy as it passes through an accelerating sample.**

Since, under the given conditions, both experiment and theory predict a quite noticeable delay in the time of flight through the crystal, it can be assumed that setting the crystal into accelerated motion should also, in accordance with (1), lead to a change in the energy of the neutrons passing through it. Such a change in energy was indeed observed in an experiment [13, 14] where neutrons with energies near the Bragg condition passed through a 5 cm thick quartz crystal oscillating in space.

It seems useful to estimate the magnitude of the time delay that occurs when a neutron passes through a crystal moving with acceleration. To make such estimates as close as possible to real experimental conditions, we used the data from [13, 14], without pretending to reproduce the hardly possible exact conditions of the experiment described therein. For the energy spectrum of neutrons, a Lorentz distribution was adopted with a center coinciding with the exact Bragg condition and a width at half maximum, $\Delta E/E \approx 3.6 \times 10^{-5}$. In this case, only neutrons from the spectrum wings could pass through the crystal. To calculate the group velocity using formulas (5)-(7) and (10), the parameters of the effective potential of the crystal given above were used. Under these conditions, a value of 11 μs was obtained for the delay in the time of flight through a crystal 5 cm thick.

Knowing the value of the time delay and the wave number of neutrons, we can estimate the value of the expected change in energy associated with the accelerated motion of the crystal. Using the data on the frequency and amplitude of the spatial oscillation of the sample, we have determined that the maximum velocity and acceleration of the sample are $v_{max}$=4.2 mm/s and $a_{max} = 1.2 \times 10^4$ cm/s², respectively. For a rough estimate of the expected effect, we will neglect the small change in the effective length of the crystal, caused by the fact of its movement, as well as the change in the speed and acceleration of the crystal during its passage by neutrons. Assuming, therefore, that during the entire time the crystal is moving with constant acceleration and taking for the neutron wavenumber the value $k_0 = 1.2 \times 10^8$ cm$^{-1}$, we obtain from formula (1) that the change in neutron energy as it passes through the sample is $\Delta E$ = 10 neV. This is consistent with the results of [13-14] within an order of magnitude.

Thus, it can be assumed that the approach based on the concept of the universal "Acceleration Effect" and formula (1) is also applicable in the case under consideration of neutrons passing through an accelerating crystal. However, with this approach, physical phenomena that lead to the slowing down or acceleration of neutrons are completely excluded from consideration.

To fill this gap, we must consistently follow the change in the wave number and frequency of the wave as it passes through the accelerating crystal. To do this we must turn again to the concept of the differential Doppler effect. Unlike the case of ordinary refraction, however, it should be taken into account that in the coordinate system of a crystal, moving with variable speed, the wave number $k_0$ also depends on time. This leads to the fact that, in accordance with the given dispersion law (4), the wave number inside the crystal also depends on time and changes as the wave propagates through the accelerating crystal. Note that this statement differs significantly from the viewpoint of the authors of the papers [13, 14].

The Doppler effect in the refraction of an electromagnetic wave at the boundary of a moving medium has been considered in [15, 16]. A similar effect for neutrons has been considered in [17]. Following mainly this work, we obtain an expression for the frequency of a neutron wave passed through an accelerating sample characterized by an arbitrary dispersion law (4). Herewith, the ratio of wave numbers in the medium and vacuum in all cases will be called the refractive index, regardless of the physical reasons leading to the difference of this value from unity and the specific type of dispersion law (4).

Representing the initial state of the neutron in the usual way as a plane wave $\psi(x,t) = \exp[i(k_0 x - \omega_0 t)]$, we find the wave function of the neutron in a moving coordinate system where the matter is at rest. According to [18] it has the form

$$\psi'(x',t) = \psi(x' + Vt, t) \exp[-i(k_V x' + \omega_V t)], \qquad (11)$$

where

$$k_V = \frac{mV}{\hbar}, \quad \omega_V = \frac{mV^2}{2\hbar}, \qquad (12)$$

and $V$ is the velocity of the medium and its boundaries at the moment the neutron enters the medium.

In this coordinate system, the wave number and frequency of the incident wave are

$$k'_0 = k_0 - k_V, \quad \omega' = \omega_0 + \omega_V - k_0 V. \qquad (13)$$

When a neutron passes through a stationary boundary of matter in this system, only the wave number changes, and the wave function inside the matter has the form

$$\psi'_i(x',t) = \exp[i(k'x' - \omega't)], \qquad (14)$$

where $k' = F(k'_0)$ is the wave number in the moving system.

Let us substitute relations (13) into (14) and then return to the laboratory coordinate system by performing the transformation

$$\psi_i(x,t) = \psi'_i(x - Vt, t) \exp[i(k_V x - \omega_V t)].$$

7Assuming that the velocity of the medium is small compared to the velocity of the neutron $v_0 = k_0 \hbar / m$, we obtain to the first order in $V/v_0$, that the frequency of the wave in the moving medium is

$$\omega_i = \omega_0 + \left(\frac{k'}{k'_0} - 1\right) k_0 V. \quad (V \ll v_0). \tag{15}$$

It is obvious that with uniform motion of the material layer, there is no final frequency change. This means that the frequency change that occurs when neutron leaving the medium differs in sign from the change that occurs when neutron entering it. Assuming this, we obtain that in the case of a sample moving with acceleration, the total frequency shift of the wave passing through the sample is determined by the expression

$$\Delta\omega = \left(\frac{k'}{k'_0} - 1\right) k_0 V - \left(\frac{k''}{k''_0} - 1\right) k_0 W, \tag{16}$$

where $W$ is the speed of the medium at the moment the neutron leaves it,

$$k''_0 = k_0 - k_W, \quad k'' = F(k''_0), \quad k_W = \frac{mW}{\hbar}. \tag{17}$$

Given the dispersion law (4) and the equation of motion of the crystal, the above relations together with formula (5) completely determine the problem. Note that in the extreme case of a small change in velocity and a weak dependence of the wavenumber on it, when the refractive index $n$ can be considered constant, formula (16) takes the form $\Delta\omega \cong k_0 (1-n) \Delta V$.

In the case of uniformly accelerated motion $\Delta V = a \Delta t$, where $\Delta t$ is the time of passage of the sample. Taking for the latter

$$\Delta t = d/(n v_0), \quad (a\Delta t \ll v_0), \tag{18}$$

where $d$ is the is the sample thickness, for the frequency change, we have:

$$\Delta\omega = k_0 \left(\frac{1-n}{n}\right) \frac{ad}{v_0}, \quad (a\Delta t \ll v_0). \tag{19}$$

Since a time delay, associated with the difference between the velocity of a neutron in a medium and in a vacuum, in this case is

$$\tau = \frac{d}{v_0}\left(\frac{1-n}{n}\right), \tag{20}$$

then formula (19) is equivalent to formula (1). As for the change in energy, it obviously is

$$\Delta E = mad\left(\frac{1-n}{n}\right), \quad (a\Delta t \ll v_0), \tag{21}$$

which is in complete agreement with the results of works [19, 20], where it was obtained by other methods.

Let us note two important circumstances here. Firstly, the above considerations are valid only in the case of the validity of the semiclassical approach to the problem, when we can talk about the moments of the neutron's entry into the moving matter and its exit into the vacuum. An important assumption here is that during the time the neutron is inside the matter, the nature of the sample's motion does not change significantly [20]. The difference in the neutron wave frequency shifts when it enters and exits the matter is also associated with the difference in the sample boundary velocities at these two time moments. Thus, we are talking about the differential Doppler effect when crossing the moving interface of the medium and vacuum. Only the velocity of the sample exit surface at the moment the neutron exits the matter depends on the specific type of dispersion function $F(k_0^2)$ and the law of sample motion $V(t)$.

Secondly, it was assumed everywhere above that the fact of acceleration of the medium does not affect the dispersion law of neutron waves in it. This assumption is not quite obviously [1].

**4. Conclusion.**

The problem of changing the energy of a neutron when it passes through an accelerating crystal under conditions close to the Bragg condition is considered. It is shown that, just as in the case of long-wave neutrons passing through a refractive sample, the accelerated motion of the sample result in a change in the neutron energy. The physical nature of the phenomenon in both cases is due to the difference in the Doppler frequency shift at the entrance and exit of the wave from the sample, which arises due to the difference in the sample velocities at the moments of entry and exit of the wave. The difference between the case of ordinary refraction of long-wave neutrons and the case of the wave passage through a crystal under conditions where diffraction has a dominant influence on the nature of wave propagation is only quantitative. Only the difference in the velocities of the input and output surfaces of the sample at different moments of their intersection by the neutron depends on the particular type of the dispersion law and the law of motion of the sample.

Approximate estimates of the magnitude of the energy change under conditions close to the conditions of the only experiment [13] are in qualitative agreement with the results of the latter. The above explanation of the nature of the phenomenon of acceleration and deceleration of neutrons when they pass through an accelerating crystal differs from the viewpoint of the authors of works [13, 14].

**References**

1. A.I. Frank, P. Geltenbort, M. Jentschel, D.V. Kustov, G.V. Kulin, V.G. Nosov, and A.N. Strepetov. Phys. At. Nucl. 71 1656‑1674 (2008).
2. A. I. Frank, P. Geltenbort, M. Jentschel, D. V. Kustov, G. V. Kulin, and A. N. Strepetov. JETP Lett. **93** (2011) 361.
3. K. Tanaka, Phys. Rev. A 25 (1982) 385.
4. A. Peres, Am. J. Phys. 51 (1983) 947.
5. A. I. Frank. Physics-Uspekhi. 53 (2020) 500.
6. C.G. Shull, A. Zeilinger, G.L. Squires, M.A. Horne, D.K. Atwood and J. Arthure. Phys. Rev. Letters 44 (1980) 1715.
7. V.V. Voronin, E.G. Lapin, S.Yu. Semenikhin and V.V. Fedorov. JETP Lett. 71 (2000) 76.
8. V.V. Voronin, Y.V. Borisov, A.V. Ivanyuta, I.A. Kuznetsov, S.Yu. Semenikhin, V.V. Fedorov. JETP Lett 96 (2013) 609.
9. A. Zeilinger, C.G. Shull, M.A. Horne, and K.D. Finkelstein. Phys. Rev. Lett. 57 (1986) 3089.
10. K. Raum, M. Koellner, A. Zeilinger, M. Arif and R. Gähler. Phys. Rev. Lett. 74 (1995) 2859.
11. A. I. Frank. Physics – Uspekhi**,** 61 (2018) 900.
12. V.K. Ignatovich. The Physics of Ultracold Neutrons (Clarendon Press, Oxford*)* 1990, p.157.
13. V.V. Voronin, Y.A. Berdnikov, A.Y. Berdnikov, Yu.P. Braginetz, E.O. Vezhlev, I.A. Kuznetsov, M.V. Lasitsa, S. Yu. Semenikhin, V. V. Fedorov. JETP Lett 100 (2014) 497.
14. Y.P. Braginetz, Y.A. Berdnikov, V.V. Fedorov, I.A. Kuznetsov, M.V. Lasitsa, S.Yu. Semenikhin, V.V. Voronin. Physica B: Condensed Mat. 551 (2017) 331.
15. C. Yeh. Journ. Appl. Phys. 36 (1965) 3515.
16. S. N. Stolyarov. Radiophys. Quantum Electron. 10 (1967) 151.
17. A.I. Frank, V.A. Naumov. Phys. At. Nucl, 76 (2013) 1423.
18. L. D. Landau and E. M. Lifshitz. Quantum Mechanics: Non-Relativistic Theory*,* 3rd edition (Pergamon press, New York) 1984, p. 52.
19. F. V. Kowalski. Phys. Lett. 182 (1993) 335.
20. V.G. Nosov and A.I. Frank. Phys. At. Nucl. **61** (1998) 613.